\documentclass[12pt]{article}

\usepackage{amsmath}
\usepackage{amssymb}
\usepackage{epsfig}
\usepackage{graphicx}
\usepackage{cite}

\makeatletter
\renewcommand{\fnum@table}{\textbf{\tablename~\thetable}}
\renewcommand{\fnum@figure}{\textbf{\figurename~\thefigure}}
\makeatother

\newcounter{myenumi}

\renewcommand{\themyenumi}{\roman{myenumi}}
{\end{list}}

\newlength{\myem}
\newcounter{mysubequation}[equation]

\makeatletter
\renewcommand{\section}{\@startsection{section}{1}{0em}{-\baselineskip}%
{\baselineskip}{\normalfont\large\bfseries}}
\renewcommand{\subsection}%
{\@startsection{subsection}{2}{0em}{-0.7\baselineskip}%
{0.7\baselineskip}{\normalfont\bfseries}}
\makeatother

\newcommand{\bi}{\begin{itemize}}
\newcommand{\ei}{\end{itemize}}

\newcommand{\be}{\begin{equation}}
\newcommand{\ee}{\end{equation}}
\newcommand{\bea}{\begin{eqnarray}}
\newcommand{\eea}{\end{eqnarray}}

\newcommand{\ldm}{\Delta m_{31}^2}
\newcommand{\sdm}{\Delta m_{21}^2}
\newcommand{\deltacp}{\delta_{\mathrm{CP}}}
\newcommand{\stheta}{\sin^2(2 \theta_{13})}

\newcommand{\ie}{{\it i.e.}}

\newcommand{\eg}{{\it e.g.}}

\newcommand{\cf}{{\it cf.}}

\newcommand{\etc}{{\it etc.}}
\newcommand{\eq}{Eq.}

\newcommand{\fig}{Fig.}

\newcommand{\Ref}{Ref.}
\newcommand{\Refs}{Refs.}

\newcommand{\equ}[1]{\eq~(\ref{equ:#1})}
\newcommand{\figu}[1]{\fig~\ref{fig:#1}}

\begin{document}

\begin{titlepage}

\renewcommand{\thefootnote}{\alph{footnote}}

\vspace*{-3.cm}
\begin{flushright}
NUHEP-TH/05-09
\end{flushright}


\renewcommand{\thefootnote}{\fnsymbol{footnote}}
\setcounter{footnote}{-1}

{\begin{center}
{\large\bf
What would it take to determine the neutrino mass hierarchy if $\theta_{13}$ were too small?
} \end{center}}
\renewcommand{\thefootnote}{\alph{footnote}}

\vspace*{.8cm}
\vspace*{.3cm}
{\begin{center} {\large{\sc
 		Andr\'e~de~Gouv\^ea\footnote[1]{\makebox[1.cm]{Email:}
                degouvea@northwestern.edu} and
                Walter~Winter\footnote[2]{\makebox[1.cm]{Email:}
                winter@ias.edu}
                }}
\end{center}}
\vspace*{0cm}
{\it
\begin{center}

\footnotemark[1]
       Department of Physics \& Astronomy, Northwestern University, \\
       2145 Sheridan Road, Evanston, IL 60208, USA

\footnotemark[2]
       School of Natural Sciences, Institute for Advanced Study, \\
       Einstein Drive, Princeton, NJ 08540, USA

\end{center}}

\vspace*{1.5cm}

{\Large \bf
\begin{center} Abstract \end{center}  }

We discuss the experimental requirements for a mass hierarchy measurement
for $\theta_{13}=0$ using muon neutrino disappearance. We find that a specially optimized neutrino factory at $L \simeq 6 \,000 \, \mathrm{km}$ could do this measurement using
extreme luminosities. In particular, we do not require charge identification
for this purpose. In order to measure the mass hierarchy for more adequate luminosities,
we explore the capabilities of low energy narrow band off-axis beams, which have relatively more events at low energies. We find that, in this case,  the energy resolution of the detector quickly becomes the limiting factor of the measurement, and significantly  affects the baseline optimization for determining the mass hierarchy.

\vspace*{.5cm}

\end{titlepage}

\newpage

\renewcommand{\thefootnote}{\arabic{footnote}}
\setcounter{footnote}{0}

\section{Introduction}

Among the goals of next-generation neutrino experiments are a measurement of the magnitude of the  $\theta_{13}$ angle of the recently revealed lepton mixing matrix (using the PDG parameterization of the mixing matrix \cite{newPDG}) and the determination of the so-called neutrino mass hierarchy.

Determining the neutrino mass hierarchy  is equivalent to establishing how the neutrino mass eigenvalues, properly defined, are ordered. It is customary to define the neutrino masses-squared $m_1^2,m_2^2,m_3^2$ such that $m_1^2<m_2^2$ and  $\Delta m^2_{21}<|\Delta m^2_{31}|$, where $\Delta m^2_{ij}\equiv m_i^2-m_j^2$. The case $m_3^2>m_2^2$ ($\Delta m^2_{31}>0$) is referred to as a normal mass hierarchy, while the other logical possibility, $m_3^2<m_1^2$ ($\Delta m^2_{31}<0$), is referred to as an inverted mass hierarchy. Current data on neutrino oscillations have allowed the determination of $\Delta m^2_{21}$ and $|\Delta m^2_{31}|$, but not the sign of $\Delta m^2_{31}$.

In the case of ``large'' $\theta_{13}$ values, several neutrino oscillation studies have been performed in order to address the optimal means for addressing the neutrino mass hierarchy. These include studies of $\nu_{\mu}\to\nu_e$ transitions in matter (see \eg\ Refs~\cite{Lipari:1999wy,Barger:2000nf,Wang:2001ys,Barger:2002xk,Huber:2002rs,Whisnant:2002fx,Minakata:2003ca,Ishitsuka:2005qi,MenaRequejo:2005hn,Albrow:2005kw,PDNOD}), $\nu_e\to\nu_{\mu}$ transitions in matter (see \eg\ \Refs~\cite{Cervera:2000kp,Barger:2000cp,Freund:2001ui,Apollonio:2002en,Huber:2002mx,Huber:2003ak,Burguet-Castell:2003vv,Albright:2004iw,Huber:2005jk}), precision studies of the neutrino flux from supernovae (see \eg\ \Refs~\cite{Lunardini:2003eh,Dighe:2003be,Bandyopadhyay:2003ts,Tomas:2004gr,Barger:2005it}), $\nu_\mu$ disappearance of atmospheric neutrinos~\cite{Bernabeu:2003yp,Palomares-Ruiz:2004tk,Gandhi:2004bj}, and combined analyses of $\nu_{\mu}$ and $\nu_e$ disappearance in vacuum~\cite{deGouvea:2005hk,Nunokawa:2005nx}. In the case of ``small'' $\theta_{13}$ values, on the other hand, the issue of determining the neutrino mass hierarchy has been neglected, until recently.

In \cite{deGouvea:2005hk}, it was shown that in the limit $\theta_{13}\to 0$, the neutrino mass hierarchy can be determined via precision studies of $\nu_{\mu}$ disappearance under rather extraordinary conditions. The discussion in \cite{deGouvea:2005hk} was, however, purely theoretical, and no attempt was made to determine whether such studies could be performed in practice. Here, we discuss the experimental requirements for establishing the neutrino mass ordering in the limit $\theta_{13}\to 0$ via long-baseline studies  of $\nu_{\mu}$ (and $\bar{\nu}_{\mu}$) disappearance.

First, we revisit in Sec. \ref{sec:physics} the analysis performed in \cite{deGouvea:2005hk} in order to discuss the conditions that need to be met in order to determine the neutrino mass hierarchy via neutrino oscillations. We review that it is imperative to combine precise measurements of the $\nu_{\mu}\to\nu_{\mu}$ survival rate at different values of $L/E$ and that at least one of the $L/E$ values should satisfy $\Delta m^2_{21}L/E\sim 1$. In Sec.~\ref{sec:requirements} we study in more detail what turns out to be the issue that more significantly limits our ability to determine the neutrino mass hierarchy --- the detector energy resolution in the case of low-energy (hundreds of MeV) muon-type neutrinos.

In Sec.~\ref{sec:conventional}, we study the ability of ``conventional'' neutrino beams and detectors to study the neutrino mass hierarchy via $\nu_{\mu}$ disappearance. After convincing ourselves that it is very unlikely that the neutrino mass hierarchy will be uncovered with these types of beams and detectors, we discuss, in Sec.~\ref{sec:pushing}, how the situation could change with improved detector energy resolution. In Sec.~\ref{sec:conclusion}, we summarize our results and conclude.

\section{Physical mechanism}
\label{sec:physics}

In the limit $\theta_{13}\to0$, which we assume applies henceforth, oscillations involving electron-type neutrinos (or antineutrinos) in the production or detection stage are sensitive only to one neutrino mass-squared difference: $\Delta m^2_{21}$. Therefore, they cannot convey any information about the sign of $\Delta m^2_{31}$, and, in order to learn about the neutrino mass hierarchy, one is forced to turn to the oscillation of muon-type neutrinos and antineutrinos.

In a nutshell, $P_{\mu\mu}$\footnote{We denote $P_{\alpha\beta}$ as the probability that a neutrino with energy $E$ produced in a weak eigenstate $\nu_{\alpha}$ is detected, having propagated a distance $L$, as a weak eigenstate $\nu_{\beta}$. $P_{\bar{\alpha}\bar{\beta}}$ is the equivalent for antineutrinos.} in vacuum is sensitive to the mass hierarchy in the following way  (for many more details, see \cite{deGouvea:2005hk, Nunokawa:2005nx}). We can express
\begin{equation}
P_{\mu\mu}=A\sin^2\left(\frac{\Delta_{21}L}{2}\right)+B\sin^2\left(\frac{\Delta_{31}L}{2}\right)+C\sin^2\left(\frac{\Delta_{32}L}{2}\right), \label{equ:Pmm}
\end{equation}
where $A,B,C$ are functions of the lepton mixing angles and $\Delta_{ij}\equiv\Delta m^2_{ij}/2E$. In the case of a normal hierarchy, $|\Delta_{31}|>|\Delta_{32}|$, while in the case of an inverted one the converse is true. Hence, if (a) one can determine that three different frequencies contribute to $P_{\mu\mu}$ and (b) $B\neq C$ (and both are known), the mass hierarchy is trivially determined. In the case of a normal mass hierarchy, the amplitude associated with the largest $|\Delta_{ij}|$ is $B$, while in the case of an inverted mass hierarchy the amplitude associated with the largest $|\Delta_{ij}|$ is $C$. Condition (a) summarizes the first challenge one needs to meet: the experimental setup should be such that $\Delta m^2_{31}$ and $\Delta m^2_{32}$ effects are distinguishable.

In practice, the situation is more involved. The reason is that $|\Delta m^2_{31}|$ is measured at the same time that the neutrino mass hierarchy is determined. It turns out that \cite{deGouvea:2005hk}, for a fixed value of $L/E$, for every $\Delta m^{2}_{31}=\Delta m^{2+}_{31}>0$, there is at least one different $\Delta m^2_{31}=\Delta m^{2-}_{31}<0$ such that
\begin{equation}
P_{\mu\mu}(\Delta m^2_{31}=\Delta m^{2+}_{31})=P_{\mu\mu}(\Delta m^2_{31}=\Delta m^{2-}_{31}).
\end{equation}
Furthermore, if $\Delta_{21}L\ll 1$,
\begin{equation}
\Delta m^{2-}_{31}=-\Delta m^{2+}_{31}+2\Delta m^2_{21}\cos^2\theta_{12},
\label{equ:xsmall_l}
\end{equation}
 independent of $L$ and $E$. The results above do not depend on whether the neutrinos are oscillating in vacuum or constant matter densities.

\begin{figure}[t]
\begin{center}
\includegraphics[width=0.8\textwidth]{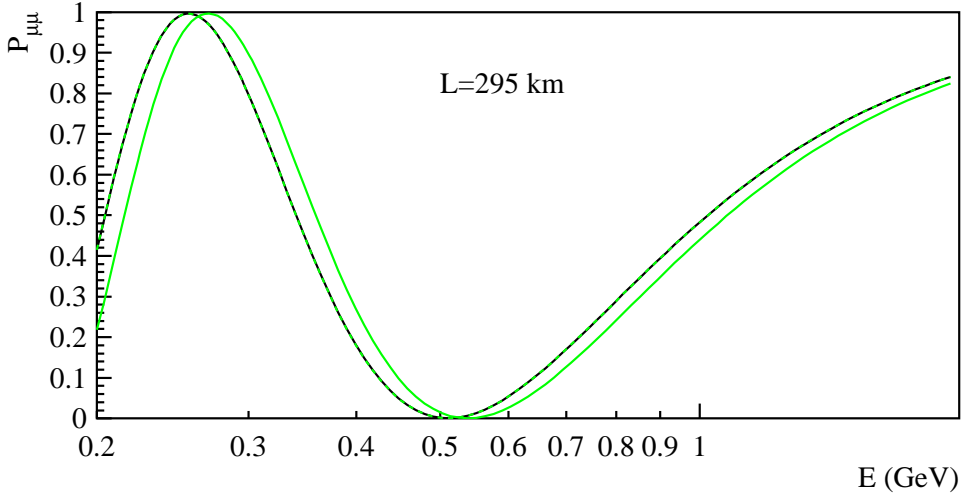}
\end{center}
\caption{\label{fig:295} $P_{\mu\mu}$ as a function of $E$ for $L=295$~km, the oscillation parameter values in \equ{oscp}, and three different values of  $\Delta m^2_{13}$: $\ldm = 2.2 \, 10^{-3} \, \mathrm{eV}^2$ (solid dark [black] curve), $\ldm = -2.2 \, 10^{-3} \, \mathrm{eV}^2$ (solid light [green] curve), and $\ldm = -2.08 \, 10^{-3} \, \mathrm{eV}^2$ (dashed light [green] curve). }
\end{figure}

 Throughout, we use the following  neutrino oscillation parameters  (\cf, \Refs~\cite{Fogli:2003th,Bahcall:2004ut,Bandyopadhyay:2004da,Ashie:2005ik,Maltoni:2004ei}), unless stated otherwise:
 \begin{eqnarray}
\sin^2 ( 2 \theta_{12})  & = &  0.83 \,  , \quad \sdm =  8.2 \, 10^{-5} \, \mathrm{eV}^2 \, ,  \nonumber \\
\stheta & = & 0 \, , \quad \sin^2 ( 2 \theta_{23})  =  1.0 \, ,  \label{equ:oscp}
 \end{eqnarray}
Note that, since we assume that $\theta_{13}$ is exactly zero, $\deltacp$ is not a physical observable.
\figu{295} depicts $P_{\mu\mu}$ as a function of $E$ for $L=295$~km, the oscillation parameter values in \equ{oscp}, and three different values of $\Delta m^2_{13}$:\footnote{According to recent analyses of atmospheric data \cite{Ashie:2005ik}, $|\Delta m^2_{31}|$ is constrained to be, at the 90\% level, between 1.9~$10^{-3}$~eV$^2$ and 3.4 $10^{-3}$~eV$^2$, while the best fit point is located between  2.0--2.5~$10^{-3}$~eV$^2$.} $\ldm = +2.2 \, 10^{-3} \, \mathrm{eV}^2$ (solid dark [black] curve), $\ldm = -2.2 \, 10^{-3} \, \mathrm{eV}^2$ (solid light [green] curve), and $\ldm = -2.08 \, 10^{-3} \, \mathrm{eV}^2$ (dashed light [green] curve). One can clearly see that there are two virtually identical curves: One associated to a normal mass hierarchy, the other to an inverted one. Furthermore, as discussed above, the two distinct $|\Delta m^2_{31}|$ values that yield the same oscillation probabilities are shifted by the amount $2\Delta m^2_{21}\cos^2\theta_{12}=1.16 \, 10^{-4} \, \mathrm{eV}^2$. A measurement of $\Delta m^2_{31}$ performed under these circumstances would yield, regardless of its precision, two degenerate values --- one associated to a normal mass hierarchy, the other associated with an inverted one. It is instructive to note that, if, for some magical reason, $|\Delta m^2_{31}|$ were independently known with great precision, it would be ``straight forward'' to tell an inverted from a normal hierarchy by measuring $P_{\mu\mu}$ at $L=295$~km (as can see by comparing the solid lines in \figu{295}).

In order to break the degeneracy, one must also study $P_{\mu\mu}$ at a different value of $L/E$. Furthermore, it is imperative that the value be ``large,'' such that $\Delta_{21}L$ is of order one, and \equ{xsmall_l} does {\sl not} apply. Numerically,
\begin{equation}
\Delta_{21} \, L=1.04 \left(\frac{\Delta m^2_{21}}{8.2 \, 10^{-5} \, \mathrm{eV^2}}\right)\left(\frac{0.6 \, \mathrm{GeV}}{E}\right)\left(\frac{L}{6 \, 000 \, \mathrm{km}}\right). \label{equ:best}
\end{equation}
Hence, $\Delta_{21}L\gtrsim 1$ implies long distances ($L\gtrsim 3\, 000$~km) and small neutrino energies, ($E\lesssim 1$~GeV).

\begin{figure}[t]
\begin{center}
\includegraphics[width=0.8\textwidth]{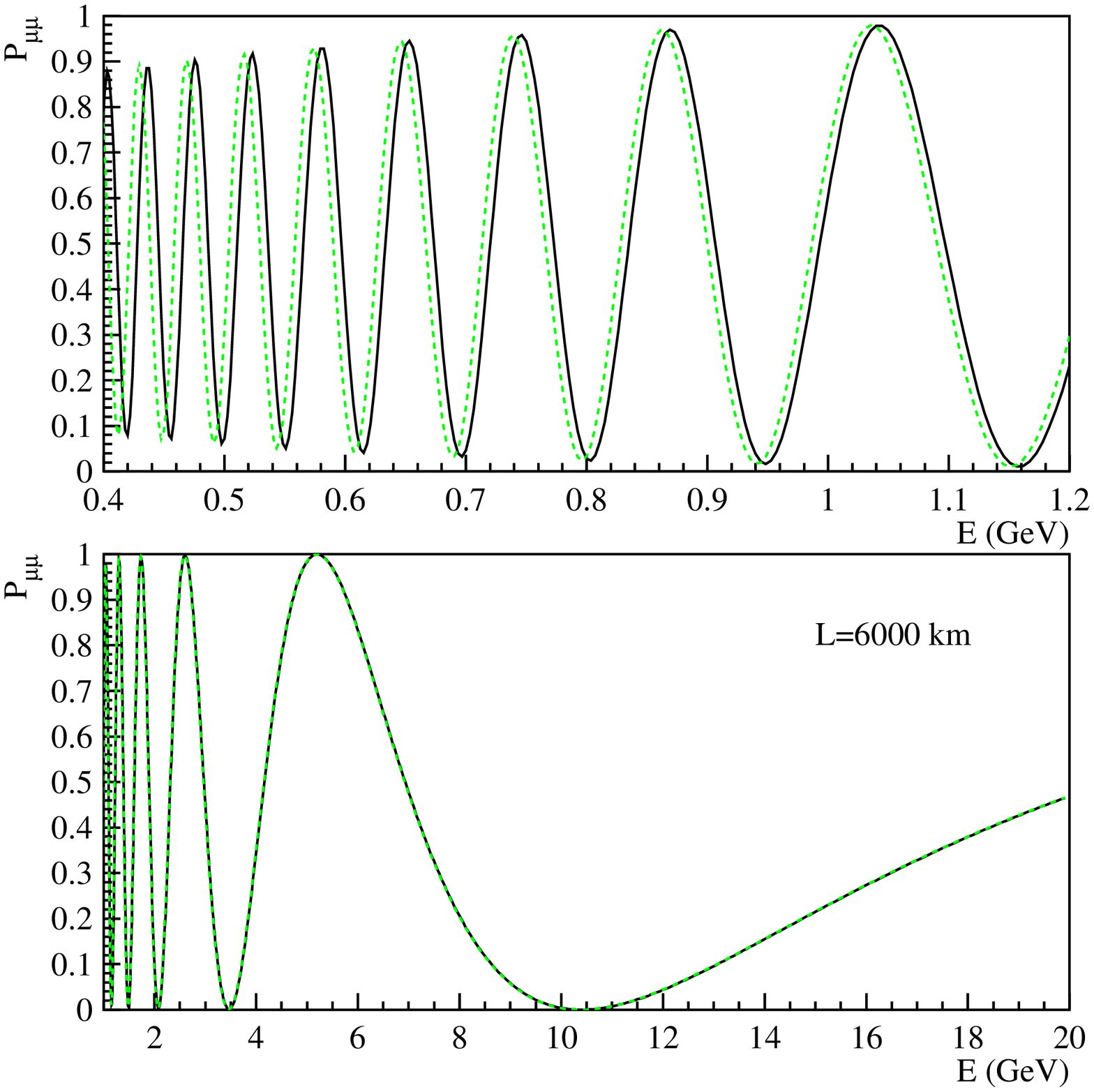}
\end{center}
\caption{\label{fig:6000} $P_{\mu\mu}$ as a function of $E$ for $L=6000$~km, the
oscillation parameters in \equ{oscp}, and $\ldm = 2.2 \, 10^{-3} \, \mathrm{eV}^2$ (solid [black] curve) and $\ldm = -2.08 \, 10^{-3} \, \mathrm{eV}^2$ (dashed [green] curve). }
\end{figure}

\figu{6000} depicts $P_{\mu\mu}$ as a function of $E$ for $L=6000$~km, the oscillation parameters in \equ{oscp}, and the two distinct values of $\Delta m^2_{31}$ that yielded the same $P_{\mu\mu}$ functions at $L=295$~km:  $\ldm = 2.2 \, 10^{-3} \, \mathrm{eV}^2$ (solid [black] curve) and $\ldm = -2.08\, 10^{-3} \, \mathrm{eV}^2$ (dashed [green] curve).
At large energies (small $\Delta_{21}L$), the situation is identical to the $L=295$~km case (\figu{295}) --- the $P_{\mu\mu}$ functions  are virtually identical when one compares the best fits to the $L=295$~km data provided by the two hypothesis concerning the mass hierarchy. The situation is quite different at low energies (order one $\Delta_{21}L$). There, the expected values of $P_{\mu\mu}$ are quite distinct, especially at very low energies.
Hence, while the first few oscillation maxima are positioned at the same energy value (e.g., in the figure these are occurring at around 10.5~GeV, 3.5~GeV, 2.1~GeV, \ldots) for both hierarchy hypothesis, higher oscillation maxima disagree. If these higher oscillation maxima were experimentally resolved, one should be able to uniquely determine the neutrino mass hierarchy. 
This is the issue on which we concentrate in the  next section. It will turn out that nontrivial information about the neutrino mass hierarchy will arrive, for $L=6000$~km, from neutrino energies $E<2$~GeV (or so). Note that while the difference between the position of the minima of the two curves is very small  for $E=1$ or 2~GeV  energies (around 10~MeV), the cumulative effect of measuring the position of several minima will prove to be important ({\it cf.} Sec.~\ref{sec:conventional}). 

\begin{figure}[t]
\begin{center}
\includegraphics[width=0.8\textwidth]{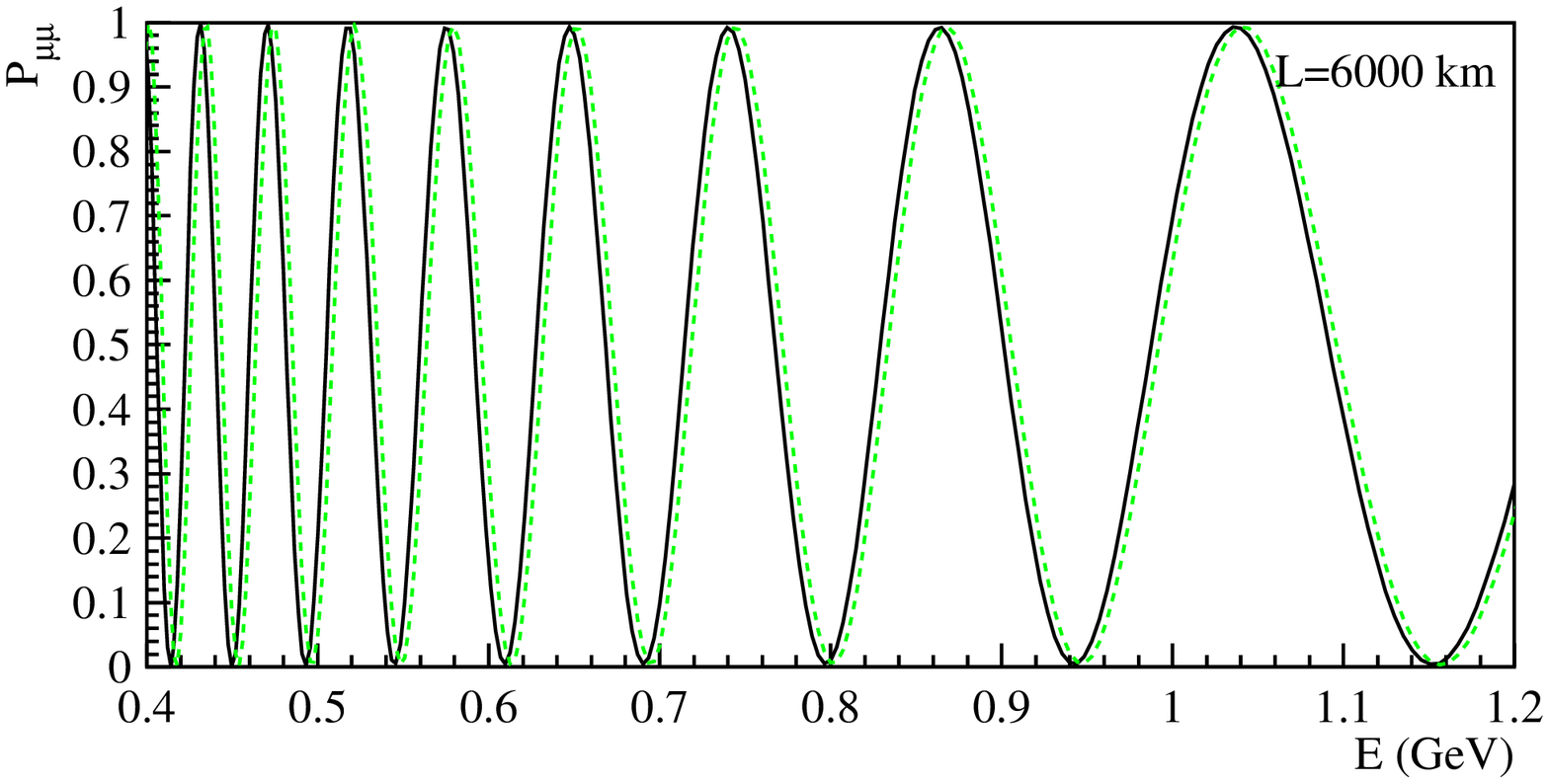}
\end{center}
\caption{\label{fig:6000a} $P_{\bar{\mu}\bar{\mu}}$ as a function of $E$ for $L=6000$~km,
the oscillation parameters from \equ{oscp}, and $\ldm = 2.2 \, 10^{-3} \, \mathrm{eV}^2$ (solid [black] curve) and $\ldm = -2.08\, 10^{-3} \, \mathrm{eV}^2$ (dashed [green] curve). }
\end{figure}

In principle, more information can be obtained if one also explores $P_{\bar{\mu}\bar{\mu}}$, the muon-type antineutrino oscillation probability. In \figu{6000a}, we depict $P_{\bar{\mu}\bar{\mu}}$ as a function of $E$ for $L=6000$~km, the oscillation parameters from \equ{oscp}, and $\ldm = 2.2 \, 10^{-3} \, \mathrm{eV}^2$ (solid [black] curve) and $\ldm = -2.08 \, 10^{-3} \, \mathrm{eV}^2$ (dashed [green] curve). As in \figu{6000} (top), the two candidate hypothesis regarding the neutrino mass hierarchy yield (in this case only very slightly) different survival probabilities as a function of energy. Most relevant here is the fact that the behavior of neutrinos and antineutrinos is distinct, which means that the ``fake solution''
patterns are shifted into different directions. Hence, if one were to combine studies of both polarities, one should be able to better discriminate the two mass hierarchies.
The origin of this distinction is due to nontrivial Earth matter effects, which are present for these very long baselines. We emphasize that these are ``solar'' matter effects, that render the effective $\theta_{12}^{\mathrm{matter}}$ and the effective oscillation frequencies $\Delta_{31}^{\mathrm{matter}}$ and $\Delta_{32}^{\mathrm{matter}}$ distinct for neutrinos and antineutrinos (for detailed expressions, see, for example, \Refs~\cite{deGouvea:2005hk,Akhmedov:2004ny}). Note that
this effect of Earth matter is conceptually different from the one used to probe the mass hierarchy via $\nu_{\mu}\leftrightarrow\nu_e$--oscillations in the case of ``large'' $\theta_{13}$ values.

\section{Experimental limitations}
\label{sec:requirements}

It is, mainly, a detector's energy resolution that determines its ability to ``see'' oscillation maxima and minima. As described above, it is this ability that will determine whether an experiment can determine the neutrino mass hierarchy via muon-type neutrino (and/or antineutrino) disappearance.

The technique used to determine the neutrino energy depends on the detector and the detection process. It is always the case, however, that the neutrino is observed via charged current scattering off some target~$X$: $\nu_{\mu} ~(\bar{\nu}_{\mu})+X\to \mu^{\mp}+X'$, and the incoming neutrino energy is reconstructed from the measured muon energy and, sometimes, the measured or inferred $X'$ energy. Water Cherenkov detectors, for example, detect neutrinos via charged-current neutrino scattering on nucleons, and the incoming neutrino energy is reconstructed by assuming that the scattering process is quasi-elastic. In calorimeter-like detectors (such as MINOS), on the other hand, the recoil hadronic energy is often also measured, and included in the reconstruction of the incoming neutrino energy.

As we are interested in the best achievable energy resolution for low-energy neutrinos, we emphasize here that, when experiments such as NOvA talk about effective energy resolutions $\Delta E / \mathrm{GeV} \sim  3\%-6\% \times \sqrt{E/\mathrm{GeV}}$,
 they refer to the relevant analysis range where their main signal is (in the case of NOvA, around a couple of GeV). This estimate would translate, for a $1 \, \mathrm{GeV}$ neutrino, into a $30 -60 \, \mathrm{MeV}$ energy resolution, and is dominated by uncertainties related to measuring the muon and recoil hadronic energies, understanding the physics of neutrino nucleon scattering in the nuclear environment, etc.  However, there is a ``physics'' lower limit on the energy resolution as $E$ decreases, dominated by the Fermi motion of  nucleons inside the nuclei. For example, the energy resolution of the QE
 events in a water Cherenkov detector is already dominated by this Fermi motion, which leads
 to an effective energy resolution of about $85 \, \mathrm{MeV}$~\cite{Itow:2001ee}.  It is in this lower limit we are ultimately interested here, and it defines what we mean by ``energy resolution'': It is the minimum achievable energy resolution for sub-GeV neutrinos.

\begin{figure}[t]
\begin{center}
\includegraphics[width=0.8\textwidth]{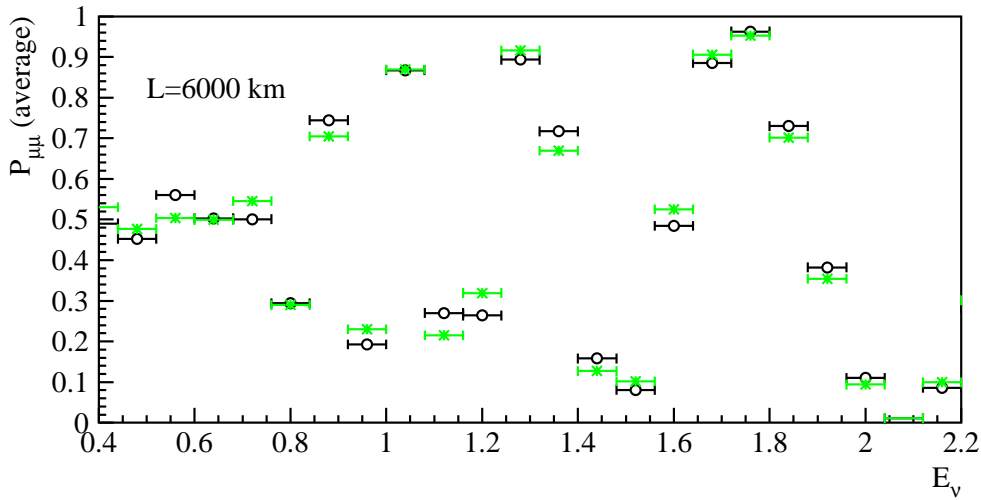}
\end{center}
\caption{\label{fig:6000average} $P_{\mu\mu}$ averaged over $\Delta E=80$~MeV energy bins, as a function of $E$ for $L=6000$~km, the oscillation parameters from \equ{oscp}, and $\ldm = 2.2 \, 10^{-3} \, \mathrm{eV}^2$ (open [black] circles) and $\ldm = -2.08 \, 10^{-3} \, \mathrm{eV}^2$ (solid [green] asterisks). }
\end{figure}
\figu{6000average} depicts $P_{\mu\mu}$ averaged over constant $\Delta E=80$~MeV energy bins
(corresponding to an energy resolution of about $\pm 40 \, \mathrm{MeV}$) as a function of $E$ for $L=6000$~km, the oscillation parameters from \equ{oscp}, and $\ldm = 2.2 \, 10^{-3} \, \mathrm{eV}^2$ (solid [black] curve) and $\ldm = -2.08 \, 10^{-3} \, \mathrm{eV}^2$ (dashed [green] curve). The averaging over energy bins is meant to illustrate the effect of a finite energy resolution: as can be seen in the figure, all ``wiggles'' below 700~MeV are erased, together with the distinction between the two different hierarchy hypothesis (compare Figs.~\ref{fig:6000average} and \ref{fig:6000}).\footnote{``Left-over'' differences between the two hierarchy hypothesis at averaged out energies are an artifact of the fixed bin size of \figu{6000average}. They are not observable in more realistic simulations, including those reported in the other sections of this paper.}

It is easy to estimate when the oscillations are ``too fast'' to be resolved. The difference between the energy of the $n$th and the $n+1$th maxima (or minima) is
\begin{equation}
\frac{E_n-E_{n+1}}{E_n}=\frac{2}{2n+3},
\end{equation}
where, in vacuum, the $n$th maximum is defined by\footnote{The distinction between oscillations in matter and in vacuum is, as far as these estimates are concerned, very small, and can be ignored.}
\begin{equation}
\frac{|\Delta m^2_{31}|L}{4E_n}=\left(2n+1\right)\frac{\pi}{2}. \label{equ:en}
\end{equation}
For a fixed energy resolution $\Delta E$, we can (from the requirement $E_n - E_{n+1} \gtrsim \Delta E$) at best hope to probe $n$ values smaller than
\begin{equation}
n_{\mathrm{max}}\sim\sqrt{\frac{|\Delta m^2_{31}|L}{4\pi\Delta E}}=8.2\left[\left(\frac{|\Delta m^2_{31}|}{2.2\times 10^{-3}~\mathrm{eV}^2}\right)\left(\frac{80~\mathrm{MeV}}{\Delta E}\right)\left(\frac{L}{6000~\mathrm{km}}\right)\right]^{\frac{1}{2}},
\label{equ:n}
\end{equation}
where we have assumed that $n_{\mathrm{max}}\gg 1$. It is clear that larger $n$ values are accessible for larger values of $L$ and smaller values of $\Delta E$. Note that
the maximum $n_{\mathrm{max}}$ is only observable if the number of events at $E_{n_{\mathrm{max}}}$ (defined in  \equ{en}) is large enough and is located within the analysis range. This means that the $n$th maximum may in fact appear below the analysis range, in which case the sensitivity of the measurement is not predominantly determined by $\Delta E$. 
Using \equ{en} and \equ{n}, we have
\begin{equation}
E_{n_{\mathrm{max}}} \propto \sqrt{\ldm \, \Delta E \, L} \, . \label{equ:enmax}
\end{equation}
Thus, smaller values of $\Delta E$ correspond to smaller
energies $E_{n_{\mathrm{max}}}$. This is easy to understand since, for fixed $L$, the ``faster'' oscillations at lower energies can be resolved with a better energy resolution (\cf, \figu{6000}). If, on the other hand, the baseline increases for fixed $\Delta E$, $n_{\mathrm{max}}$ will increase according to \equ{n}
and will move to higher energies according to \equ{enmax}, because the oscillations become faster
at the energy where it was before.
In practice, statistics plays a crucial role in
realistic experiments. For instance, if we assume a very narrow band beam which only has events
close to $E$, then the constraint $E \simeq E_{n_{\mathrm{max}}}$ implies with \equ{enmax} that the optimum baseline $L$ will scale as $L \propto (\ldm \, \Delta E)^{-1}$ to resolve $n_{\mathrm{max}}$. On the other hand, the $1/L^2$ drop of the flux and the shape of the spectrum might actually imply that it is more efficient to resolve $n_{\mathrm{max}}-1$ at a shorter baseline (\cf, \equ{n}). In the next two sections, we will see that it is a non-trivial interaction between these factors which determines the baseline optimization for narrow band beams.

 In summary, while the effect we are interested in is concentrated at the lowest energies, the finite detector energy resolution washes out the effect, and requires one to obtain information regarding the mass hierarchy from intermediate energies. In order to understand the consequences of the interplay between these two conflicting requirements, detailed simulations are required. These are described in detailed in the next sections.

\section{Conventional methods}
\label{sec:conventional}
In this section, we focus on established beam and detector technologies, such as detector
technologies currently used or in active preparation. In addition, we
do not assume an energy resolution pushed to the extremes. We first discuss
how one would design such experiments with respect to the mass hierarchy
measurement, and then evaluate their performance.

\subsection{Description of experiments}

\begin{figure}[t]
\begin{center}
\includegraphics[width=\textwidth]{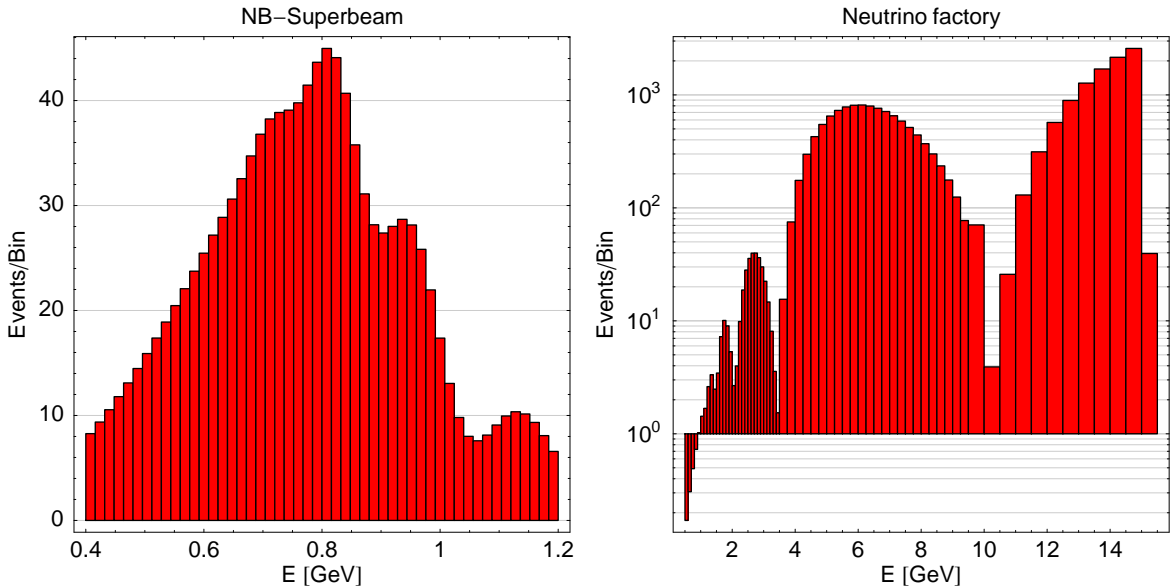}
\end{center}
\caption{\label{fig:spectra} The binned energy spectrum for the disappearances channels of
our NB-Superbeam (left, $L= 6 \, 600 \, \mathrm{km}$) and
Neutrino Factory (right, $L=6 \, 000 \, \mathrm{km}$), where the energy resolution of the NB-Superbeam has been exaggerated (improved) by a factor of two to identify the wiggles.
For the oscillation parameters, we use the values in \equ{oscp}, and $\Delta m^2_{31}=+2.2 \,10^{-3}$~eV$^2$.}
\end{figure}

Following the results of the previous sections, we will concentrate on setups capable of measuring $\ldm$ via $\nu_{\mu}$ and/or $\bar{\nu}_{\mu}$ disappearance with high precision (high luminosities) at (at least)
two different values of $L/E$ and at small energies $E \lesssim 1 \, \mathrm{GeV}$.
In particular, we are interested in the ability of such experiments to  resolve high oscillation maxima,
which means that they required very good energy resolution. We consider
two possibilities for this measurement at two distinct values of $L/E$:
\begin{enumerate}
\item
 Two Narrow Band (NB) beams at different values of $L$.
\item
 One Broad Band (BB) beam with a wide range in $E$.
\end{enumerate}
In order to probe option (1), we assume a well-established detector and beam, namely, we combine an off-axis superbeam with a water Cherenkov detector, using the T2HK proposal as an archetype.
This experiment peaks at an energy of about $0.76 \, \mathrm{GeV}$, which is perfectly suited for
our purposes. To study option (2), one could use either a broad band superbeam or a neutrino
factory beam with much higher statistics. Here, 
we concentrate on a neutrino factory beam in order to test higher statistics signals. Note that, in principle,
a broad band superbeam could replace two narrow band beams, but the detector requirements
will be similar and the events in the relevant (low) energy range will rather decrease than
increase. One potential candidate for such a broad band superbeam is the BNL-Homestake beam~\cite{Beavis:2002ye}.

\subsubsection*{Narrow band superbeam and water Cherenkov detector}

For the narrow band superbeam (NB-Superbeam), we use a setup similar to the
T2HK off-axis superbeam~\cite{Itow:2001ee} with the experiment simulation
from \Ref~\cite{Huber:2002mx} using the GLoBES software~\cite{Huber:2004ka}.
This superbeam uses a target power of $4 \, \mathrm{MW}$, a baseline of $295 \, \mathrm{km}$
in its T2HK configuration, and a water Cherenkov detector with a fiducial mass of $1 \, \mathrm{Mton}$. For the running time and polarity, we choose eight years of neutrino running
only unless stated otherwise. The energy resolution for the QE events is mainly determined by
the Fermi-motion of the nucleons in the oxygen core, which means that one has an effective
energy resolution of about $\sigma_E \simeq 85 \, \mathrm{MeV}$.\footnote{For the experiment simulation, we use an energy resolution defined as a Gaussian mapping from the incident neutrino energy $E$ to the reconstructed neutrino energy $E'$: $R(E,E') = (\sigma(E) \sqrt{2 \pi})^{-1} \, \exp \left( -\frac{(E-E')^2}{2 \sigma^2(E)} \right)$~\cite{Huber:2004ka}, where
$\sigma(E)$ is the energy resolution $\Delta E$ as function of $E$.} Note that this
energy resolution is approximately constant in energy. Compared to \Ref~\cite{Huber:2002mx},
we change the binning and sampling points\footnote{Number of points, where the oscillation
probabilities are evaluated.} in order to be able to resolve fast oscillations and higher oscillation maxima. We use 50 (equally spaced) bins between $0.4$ and $1.2 \, \mathrm{GeV}$ and 75 (equally spaced) sampling points between $0.2$ and $1.4 \, \mathrm{GeV}$ to avoid aliasing effects from the energy resolution function.\footnote{Since the event rates at the lower and upper end of the spectrum are not exactly zero, events beyond the analysis range could
contribute via the energy resolution function. Therefore, it is advisable to use a larger
sampling range than bin range.} This corresponds to a constant bin/sampling width of
$\Delta E = 16 \, \mathrm{MeV}$, which means that, in principle, energy resolutions up to
$\Delta E = 8 \, \mathrm{MeV}$ can be tested. However, this sampling width is not enough to
calculate very fast oscillations at low energies. Therefore, we
use the low-pass filter of GLoBES (\cf, GLoBES manual) to average out fast oscillations already at the probability level, and average over the bin width $\Delta E= 16 \, \mathrm{MeV}$.
Since we have now artificially increased our effective energy resolution by averaging at
two different places, we subtract this probability-level averaging effect from the energy resolution $\sigma_E$ such that we obtain an effective energy resolution $\sigma_{\mathrm{eff}} \simeq \sqrt{\sigma_E^2 + (16 \, \mathrm{MeV})^2} \simeq 85 \, \mathrm{MeV}$ (energy resolution type~2 in GLoBES). In \figu{spectra} (left), we show
a typical energy spectrum with an energy resolution $\sigma_{\mathrm{eff}}$ exaggerated
by a factor of two (the bin-widths are $16 \, \mathrm{MeV}$, considerably smaller than the energy resolution). One can see that there are enough events/bin to contain significant statistical information, while the ``wiggles'' can be resolved for $E\gtrsim 700$~MeV.

\subsubsection*{Wide band neutrino factory beam and TASD}

We choose a neutrino factory similar to \Ref~\cite{Huber:2002mx} with $1.06 \cdot 10^{21}$ useful muon decays per year and a total running time of four years in each polarity (corresponding
to $5.3 \cdot 10^{20}$ useful muon decays per year and polarity for a simultaneous operation with both polarities). Since we want to have as many events as possible at low energies,
we use a lower muon energy of $E_\mu=15 \, \mathrm{GeV}$. Though the total event rate
for this choice is much lower than for more ``canonical'' $E_\mu \sim 20 - 50 \, \mathrm{GeV}$, it turns out that
the relative increase in the low energy events is favorable for the performance. For the
detector, we use the Totally Active Scintillator Detector (TASD) from \Ref~\cite{Huber:2005jk} with a fiducial mass of $50 \, \mathrm{kt}$, which is similar to the NOvA detector~\cite{Ayres:2004js}.
In principle, one could also use an iron calorimeter, but the energy resolution and
efficiency of the TASD is much higher. Finally, we do not require charge
identification, as it would drastically reduce the efficiencies at low energies.
Instead, we effectively measure the disappearance rates
$N^{\mathrm{eff}}_{\mu \mu} = N_{\mu \mu} + N_{\bar{e} \bar{\mu}}$ or $N^{\mathrm{eff}}_{\bar{\mu} \bar{\mu}} = N_{\bar{\mu} \bar{\mu}} + N_{e \mu}$,
where the event rates $N_{\alpha \beta}$ correspond to $P_{\alpha \beta}$ convoluted with
the corresponding beam spectrum, cross sections, etc. This implies that the detector cannot distinguish muons from antimuons.

For the binning, we use a somewhat more sophisticated approach compared to the superbeam one described above,
because the neutrino factory spans a very large energy range with oscillation probabilities changing slowly in the high energy range and very quickly in the low energy range. We
divide the energy analysis range from $0.5$ to $15 \, \mathrm{GeV}$ into variable bins
with sizes of $100 \, \mathrm{MeV}$ until $3.5 \, \mathrm{GeV}$, $250 \, \mathrm{MeV}$ until $9.5 \, \mathrm{GeV}$, and $500 \, \mathrm{MeV}$ for the rest, where we include an additional
bin over $15 \, \mathrm{GeV}$ to reduce aliasing effects. For the muon energy resolution, we use
$0.03 \, \sqrt{E/\rm GeV}$. Since this corresponds to an energy resolution of only $21 \, \mathrm{MeV}$ at the lower end of the analysis range and the Fermi motion contribution is much higher, we add an estimated resolution of $\Delta E = 85 \, \mathrm{MeV}$ via the filter feature of
GLoBES (carbon should have a similar order Fermi motion contribution compared to oxygen). This yields an effective best energy resolution $\sigma_{\mathrm{eff}} \simeq \sqrt{(85 \, \mathrm{MeV})^2 + (16 \, \mathrm{MeV})^2} \simeq 87 \, \mathrm{MeV}$ at the lower end of the spectrum, and $\sigma_{\mathrm{eff}} \simeq 100 \, \mathrm{MeV}$ at the energy range where first significant contributions to the event rates come in. We show in \figu{spectra} (right),
a typical energy spectrum for the neutrino factory. At low energies, the bins are chosen
fine enough to resolve fast oscillation not averaged out by the energy resolution, but not too fine such that there are enough events to significantly resolve the left-most oscillation maxima.
As for most broad band beams, the lower energies are suppressed by the beam spectrum.
This implies that though the mass hierarchy effects become smaller for larger energies, the statistics rapidly increases and partially compensates for this decrease. Thus, compared to \figu{spectra} (left), the broad band beam's performance should be significantly determined by more than one oscillation maximum.

\subsection{Performance analysis}

Before we come to the analysis of the described experiments, we need to define our
performance indicator:
We define that we have sensitivity to the normal (inverted) mass hierarchy, if there
is no solution with the inverted (normal) mass hierarchy fitting the original solution
below the chosen confidence level. Therefore, the mass hierarchy sensitivity is
in practice determined by the $\mathrm{sgn}(\ldm)$-degeneracy~\cite{Minakata:2001qm}.
Since it turns out that for the disappearance channel in the limit $\stheta=0$ the
position of this degeneracy is mainly determined by a different value of $\ldm$ \cite{deGouvea:2005hk},
we locate it in $\ldm$-direction first and then marginalize over all oscillation parameters
(including $\ldm$) using this starting point. In the marginalization, we assume
an external precision of 5\% for $\sdm$ and 10\% for $\theta_{12}$, which is a
conservative choice for the time when the experiments are being analyzed
(\cf, \eg, \Ref~\cite{Bahcall:2004ut}). Furthermore, we
include a matter density uncertainty of 5~\%~\cite{Geller:2001ix,Ohlsson:2003ip,Pana}.
We do not include external constraints for $\stheta$.

\begin{figure}[t]
\begin{center}
\includegraphics[width=\textwidth]{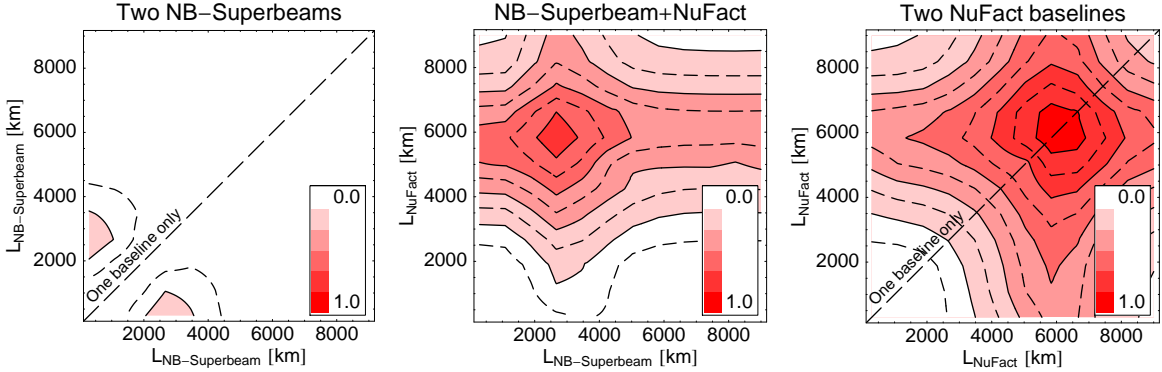}
\end{center}
\caption{\label{fig:twobase}
$\Delta \chi^2$ for the test of the normal hierarchy
as function of two baselines for the different experiment combinations
as given in the plot labels. The different contour curves correspond to $\Delta \chi^2 = 0.05$ (dashed), $0.1$ (solid), $0.15$ (dashed), $0.2$ (solid) \etc , whereas the different
shadings correspond to $\Delta \chi^2$ increased in steps of $0.1$. The diagonal lines
represent a single baseline only, \ie, all detector mass put to one baseline. For the
oscillation parameters, the values in \equ{oscp} are used, and $\Delta m^2_{31}=+2.2 \, 10^{-3}$~eV$^2$.}
\end{figure}

Since we know that, in principle, we need at least two different values of $L/E$ to resolve the
neutrino mass hierarchy, we combine {\em a priori} two different experiments/baselines to
demonstrate the effects. In \figu{twobase}, we show the $\Delta \chi^2$
for the test of the normal hierarchy as function of two baselines for the different experiment combinations as given in the plot labels. We first note that the absolute $\Delta \chi^2$
in these figures is at maximum one, which means that the experiments considered are not sufficient
for the measurements and require substantial luminosity upgrades. However, these figures are useful to discuss the optimization of these experiments for the mass hierarchy measurement.

For the combination of two NB-Superbeam  baselines (left panel of \figu{twobase}), we find the optimal performance for the combination $295 \, \mathrm{km}$ (lower end of plot) + $2 \, 700 \, \mathrm{km}$. This is what one would expect from a narrow band beam: One baseline is not sufficient to resolve the mass hierarchy. Instead, one needs the precise $\ldm$ measurement at two very distinctive values of $L/E$. It is not very surprising that the baseline $L = 295 \, \mathrm{km}$ is good for
a precision measurement of $\ldm$ because it corresponds to the original T2HK setup. This baseline
basically determines the value of the ``fake'' $\ldm$ of the inverted sign solution, which is then compared to the ``fake'' $\ldm$ at the longer baseline where solar effects contribute.
The inconsistency between these two $\ldm$'s is then represented by the $\Delta \chi^2$.
As we will see later, the length of the long baseline depends  on the
energy resolution. It is determined by a balance between several competing factors. As discussed in Sec.~\ref{sec:requirements}, the energy resolution (and the beam energy profile) determines the smallest energy where one can still ``measure'' $\Delta m^2_{13}$. With that in mind, the optimum baseline is such that the largest possible oscillation maximum can be resolved with enough luminosity (larger $n_{\mathrm{max}}$ prefers larger values of $L$, where more statistical significance is obtained than for shorter baselines;
\cf, \equ{n}).

As already mentioned above, the luminosity is still not at all sufficient for a significant result: In this case,
a factor of $30$ luminosity upgrade would be necessary for a $3 \sigma$ signal.
This corresponds to a total
luminosity of $\sim 480 \, \mathrm{Mton} \, \mathrm{yr}$ ($4 \, \mathrm{MW}$ beam power), which is not achievable in practice. A somewhat better result can be achieved by using neutrinos and antineutrinos ($2 \, \mathrm{yr}+6 \, \mathrm{yr}$) at one baseline $L \simeq 2 \, 700 \, \mathrm{km}$  instead of two different baselines. In this case, a factor of $20$ luminosity upgrade would be sufficient. In summary, two narrow band superbeams do not seem to have sufficient statistics for this measurement, but the theoretical effect is, in principle, observable.

\begin{figure}[t]
\begin{center}
\includegraphics[width=\textwidth]{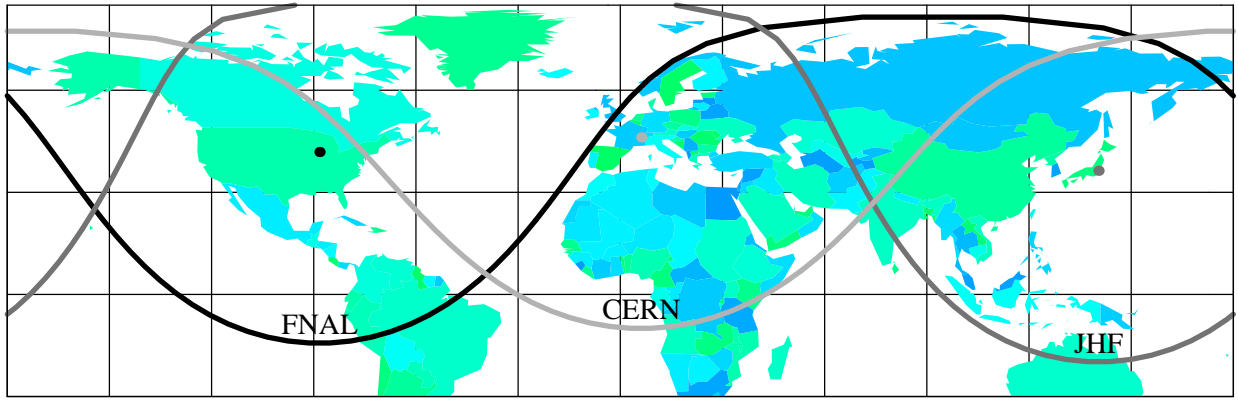}
\end{center}
\caption{\label{fig:baseplot} Potential detector locations $L=6 \, 000 \, \mathrm{km}$ for three different potential neutrino factory laboratories.}
\end{figure}

The ideal instrument to boost the statistics of such a measurement is a neutrino factory.
\figu{twobase} (middle) depicts the reach of combining data from the NB-Superbeam and the neutrino factory
we introduced above. A neutrino factory at $L \sim 6 \, 000 \, \mathrm{km}$ would clearly enhance the potential and complement
the $L \simeq 2 \, 700 \, \mathrm{km}$ baseline of the NB-Superbeam. Note that the $\Delta \chi^2$ of the combination of these two experiments is higher that the sum of the $\Delta \chi^2$'s of the neutrino factory alone and the superbeam combination alone, which means that there is real synergy between these two experiments beyond just adding their statistics. For
two neutrino factories (case depicted in \figu{twobase} (right)), it is eventually best to put all
detector mass at $L \sim 6 \, 000 \, \mathrm{km}$. As one may expect, the broad band nature of the beam itself has enough oscillation maxima to access the mass hierarchy, as long as the baseline is long enough. In this case, $\Delta \chi^2 > 1$ can be achieved, and one would require $6.8 \, \mathrm{Mton} \, \mathrm{yr}$ of data for a $3 \sigma$ signal with the TASD. Thus, given a detector ten
 times as big as NOvA, one could determine the neutrino mass hierarchy via $\nu_{\mu}$ disappearance after 23 years of data taking, but without
pushing the technical limits to the extreme. 

One could also use a NO$\nu$A-like superbeam experimental setup~\cite{Ayres:2004js} and obtain similar quantitative results. Since the NO$\nu$A  superbeam has an energy spectrum peaking around $\sim 2 \, \mathrm{GeV}$, one can read off from
 \figu{spectra} (right) that this experiment could provide the necessary information about the higher oscillation maxima, too. For example, the required luminosity for the NuMI beam with the Fermilab proton driver (see \Refs~\cite{Albrow:2005kw,PDNOD} for the physics potential) would
 be around $19 \, \mathrm{Mton} \, \mathrm{yr}$ (for $3\sigma$) using a comparable detector technology to the neutrino factory in the same off-axis configuration as NO$\nu$A. In this case, the optimal baseline would be around $5 \, 500 \, \mathrm{km}$ (similar to the neutrino factory), and neither the performance nor the baseline optimization would be significantly affected by a better energy resolution. Because of the conceptual similarity to the neutrino factory, we do not follow this approach anymore, and use ``NB-Superbeam'' to refer to a low energy ($E \lesssim 1 \, \mathrm{GeV}$) narrow band beam.

Though we have found that a lot of luminosity is required, the final increment required after a long fruitless search for nonzero $\stheta$ effects may not be completely unrealistic. As depicted in \figu{baseplot}, there are many potential detector locations for such a long baseline and different potential neutrino factory laboratories.
For example, detectors could be located in the Soudan mine for neutrinos coming from CERN or RAL, in India
from JHF or CERN, in Finland from FNAL or BNL, in the Gran Sasso laboratory from BNL, or in
Australia from JHF.

\begin{figure}[t]
\begin{center}
\includegraphics[width=0.5\textwidth]{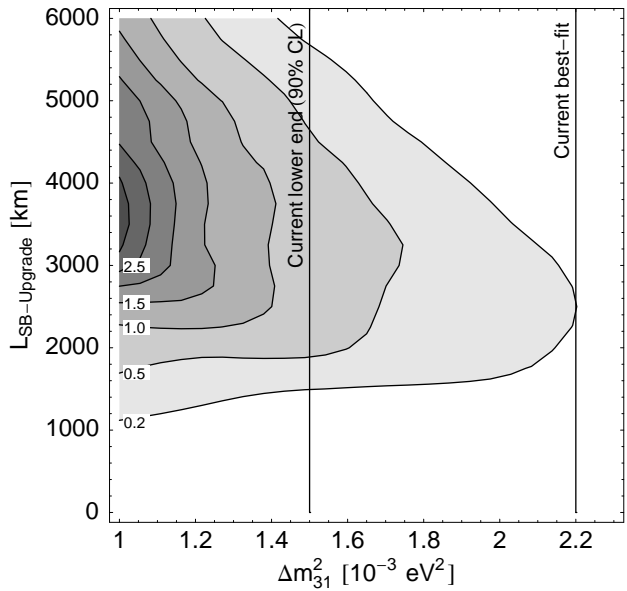}
\end{center}
\caption{\label{fig:dmdep}
$\Delta \chi^2$ for the test of the normal hierarchy
as function of the simulated $\Delta m_{31}^2$ and baseline $L$ of the SB-Upgrade, where the SB-Upgrade is operated with two years of neutrino running and six years of antineutrino
running and not combined with any other experiment. The vertical lines correspond to
the current best-fit value (right) and lower end of the allowed range (left, 90\% CL~\cite{Ashie:2005ik}).
The numbers give the $\Delta \chi^2$ for the respective contours, where it is increased
in steps of $0.5$ (except first contour).
}
\end{figure}

We now discuss the robustness of these results. As we will see in the next section,
the energy resolution of the superbeam detector can drastically change its potential to determine the mass hierarchy, as well as the optimum baseline for doing the job. Hence, as already advertised, the detector performance is the ``bottleneck.'' For the neutrino factory, on the other hand, it is the small number of events  at low energies that limits the
performance: One can see  in \figu{spectra} (right) that a factor of two better or worse  energy resolution does not change qualitatively the potential to resolve the highest significantly contributing
oscillation maximum. For $\Delta E \gtrsim 250 \, \mathrm{MeV}$, however,
we find a substantial degradation of the mass hierarchy sensitivity (larger than 50\% change in
 $\Delta \chi^2$). In addition, the energy resolution hardly affects the baseline
 optimization of the neutrino factory.

 Another important issue is the precise knowledge
 of the oscillation parameters other than $\ldm$. We find that perfect
 knowledge of these parameters would translate into a marginal 20-30\% reduction the luminosity requirements for all setups considered above.
Finally, the impact of the hierarchy (sensitivity to normal or inverted hierarchy)
turns out to be rather small and not to affect the optimization discussion.
In summary, the energy resolution is the limiting factor for the NB-Superbeam,
while the lack of statistics at low energies limits the ability of  the neutrino factories to uncover the neutrino mass ordering in the limit $\theta_{13}\to 0$.

Before proceeding, it is interesting to comment on the dependence of the sensitivity to the mass hierarchy on the simulated value of $\ldm$. \figu{dmdep} depicts the $\Delta \chi^2$ for the test of the normal hierarchy
as function of the simulated $\Delta m_{31}^2$ and baseline $L$ of the NB-Superbeam. In this case,
we choose an option with two years of neutrino running and six years of antineutrino
running not combined with any other experiment (one experiment only). As discussed above,
the $\Delta \chi^2$ at the current best-fit value of $\ldm$ requires substantial luminosity
upgrades. If, however, $\ldm$ was at the lower end of the currently allowed region, a
factor of four higher $\Delta \chi^2$ can be obtained at a somewhat longer baseline.
For even smaller values of $\ldm$, even a 90\% CL hint would be possible with this
experiment only, where the optimal baseline increases with decreasing $\ldm$. The
reason for this better performance is that the solar and atmospheric mass-squared differences
are getting closer for smaller values of $|\ldm|$, and thus any effect which requires
the presence of both oscillations becomes enhanced.
The longer baseline becomes necessary
because of the fixed peak of the energy spectrum where oscillations can be resolved with the
chosen energy resolution the relationship $\ldm \times L \sim$~const. (weighted with
$1/L^2$ event rate drop) applies (\cf, \equ{enmax}). On the other end of \figu{dmdep}, we find
a substantial loss of sensitivity above $\ldm \gg 2.5 \cdot 10^{-3} \, \mathrm{eV}^2$, which means that the mass hierarchy measurement with a NB-Superbeam becomes very hard if $\ldm$ turns out to be much larger than the current best-fit value. In this case only a neutrino factory approach
could be possible, as it turns out that the neutrino factory performance and optimization
is rather unaffected by the simulated value of $\ldm$ in the currently allowed $\ldm$-range
(90\% CL).

\section{Pushing the limits}
\label{sec:pushing}
\begin{figure}[t]
\begin{center}
\includegraphics[width=\textwidth]{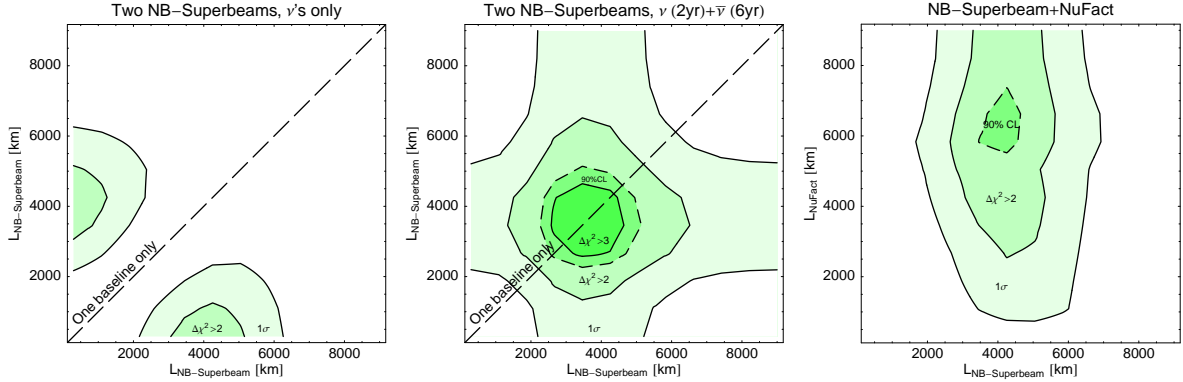}
\end{center}
\caption{\label{fig:respushed}
$\Delta \chi^2$ for the test of the normal hierarchy
as function of two baselines for the different experiment combinations
as given in the plot labels. In this figure, the energy resolution of the SB-Upgrade
is pushed by a factor of two beyond the current understanding. For the
oscillation parameters, the values in \equ{oscp} are used, and $\ldm = +2.2 \, 10^{-3}$~eV$^2$.
}
\end{figure}

From the last section, it is clear that the neutrino factory capability to determine the mass hierarchy in the limit $\theta_{13}\to0$ is hard to improve
because of the limited statistics at low energies. One could, of course,  always
think about making use of an even lower energy neutrino factory. We, however, regard this as qualitatively similar to a
narrow band beam peaking at lower energies and therefore choose the NB-Superbeam as a representative for a beam with good low-energy statistics. In this section, we discuss the consequences of  pushing the detector technology beyond ``current'' limitations.

\figu{respushed} depicts the combination of two NB-Superbeams (left -- eight years neutrino running only), two NB-Superbeams (middle -- two years $\nu$ running and six years $\bar{\nu}$ running), and a NB-Superbeam with the neutrino factory, as defined in the previous section (right). In this figure,
the maximum energy resolution of the NB-Superbeam(s) was reduced by a factor of two with respect to
the Fermi motion-dominated effective energy resolution of about $85 \, \mathrm{MeV}$.
Comparing \figu{respushed} to \figu{twobase}, we can read off that  the baseline optimization for the longer NB-Superbeam baseline changes. In this case, the optimum is at  $L \simeq 4 \, 000 \, \mathrm{km}$, longer than the value obtained with worse energy resolution depicted in  \figu{twobase}.
The reason for the increase of the optimal baseline with the improved energy resolution, clearly observed here, was discussed in Sec.~\ref{sec:requirements}.

Instead of using two NB-Superbeams at different baselines, a NB-Superbeam  experiment with a single baseline which takes advantaged of  combined neutrino and antineutrino running can exceed the
90\%CL threshold, as depicted in \figu{respushed} (middle). Similarly, the combination with a neutrino factory instead of a second superbeam
would yield even better results because of the high statistics of the neutrino factory (see
\figu{respushed} (right)). Note that \figu{respushed} (right) is ``orthogonal'' to \figu{twobase} (middle), which means that, here, the neutrino factory plays a secondary role when it comes to determining the mass hierarchy (even though one NB-Superbeam alone is not capable of providing a 90\%CL signal). By comparing the left and right
panels of \figu{respushed}, we find that, say, a shorter baseline $L \simeq 3 \, 000 \, \mathrm{km}$ neutrino factory (plus the appropriate detector) could replace the role of the shorter NB-Superbeam baseline in order to yield similar results. Such a setup may already be available by the time the measurement  discussed here is contemplated.  In summary, we learn from \figu{respushed} that an improved energy resolution would
make the NB-Superbeam the key experiment for the mass hierarchy measurement for $\stheta=0$,
where the optimal long baseline hardly depends on the combination of experiments considered,
but depends more strongly on the achievable energy resolution.

\begin{figure}[t]
\begin{center}
\includegraphics[width=0.5\textwidth]{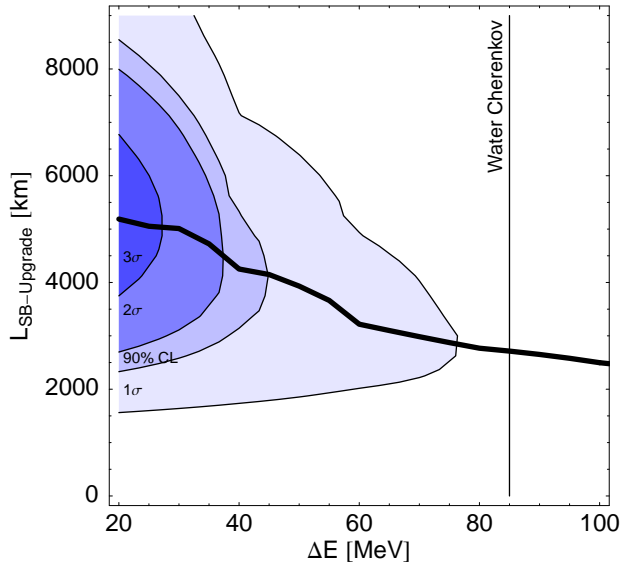}
\end{center}
\caption{\label{fig:resdep}
$\Delta \chi^2$ for the test of the normal hierarchy
as function of the energy resolution
$\Delta E$ and baseline $L$ of the SB-Upgrade, where the SB-Upgrade is combined with
a neutrino factory at $6 \, 000 \, \mathrm{km}$. The thick curve represents the
optimum baseline for each individual energy resolution. The vertical line corresponds to
the Water Cherenkov detector's effective energy resolution dominated by Fermi motion. For the
oscillation parameters, the values in \equ{oscp} are used, and $\ldm = +2.2 \, 10^{-3}$~eV$^2$.
}
\end{figure}

How good an energy resolution do we really need in order to significantly
boost the capabilities of a NB-Superbeam to determine the neutrino mass hierarchy in the limit $\theta_{13}\to 0$? \figu{resdep} depicts $\Delta \chi^2$ as function of the energy resolution $\Delta E$ and the baseline $L$ of the SB-Upgrade, where the SB-Upgrade is combined with a neutrino factory at $6 \, 000 \, \mathrm{km}$. From this
figure, one can read off a number of different results: First, the optimal baseline
of the NB-Superbeam becomes longer as the energy resolution improves (thick curve).  Second, for the energy resolution of the Water Cherenkov detector (vertical line), there is no statistically significant signal and tremendous luminosity upgrades are required. However, for better energy resolution, the statistical significance is markedly improved. A factor of four better energy resolution would allow one to obtain a $3 \sigma$ determination of the mass hierarchy for realistic luminosity levels.
And third, the gradient in the direction of the energy resolution is rather significant, which means that any improvement of the energy resolution would reduce the required
luminosity for a $3 \sigma$ signal. Note that though \figu{resdep} is computed for the
combination of the NB-Superbeam with the neutrino factory, these results are qualitatively unaffected by the particular combination of experiments (assuming it includes a NB-Superbeam).

How would one go about improving the energy resolution? A detailed study of this issue is beyond the intentions of this paper, but we wish to add a few qualitative remarks. As already alluded to above, the source of energy resolution boils down, at the levels in which we are interested, to the Fermi motion of the nucleons inside the nuclei that make up the target. In a water Cherenkov detector, for example, sub-GeV neutrinos scatter, mostly quasi-elastically, off of nucleons contained in oxygen nuclei. The Fermi motion of these leads to the intrinsic $\Delta E\sim 85$~MeV quoted above. There are, at least in principle, a few ways to improve on this. Different materials have different intrinsic uncertainties on the nucleon energies. One naively expects that materials with larger mass number ($A$) would have smaller intrinsic $\Delta E$ values. One the other hand, detectors that are capable of fully reconstructing the final state hadronic state in an event-by-event basis (in the case of quasi-elastic events, fully determining the energy and momentum of the final state daughter nucleon) may bypass this uncertainty. Furthermore, if one could afford a pure proton (or even electron!) target, Fermi motion would be absent (replaced, perhaps, by much less significant thermal motion) and one could hope to improve on the energy resolution. Finally, energy resolution issues could be effectively erased if mono-energetic muon-neutrino beams were available (such as neutrino beams from muon capture on nuclei, similar to \cite{Sato:2005ma},  or neutrino beams from stopped pion decay). In this case, one could repeat the analysis done here by placing different experiments at several different baselines in order to measure $\Delta m^2_{31}$ as a function of $L/E$.

While some of these propositions sound intriguing, we remind readers that in order to perform the measurement we are proposing, extreme luminosities are required: not only do we need a more precise detector technology, but we also need one that is capable of accumulating data as quickly as the  ``bread-and-butter'' designs  which are currently available. Our hope here is not to resolve this issue in any way, but to attract the attention of the community to these matters.

\section{Summary and conclusions}
\label{sec:conclusion}

Determining the neutrino mass hierarchy is of the utmost importance, but may prove very challenging if $\theta_{13}$ is too small. In the near future, we hope to learn, through next-generation long-baseline and reactor experiments, whether $\sin^22\theta_{13}\gtrsim 10^{-2}$ (see, for instance, \Ref~\cite{Huber:2004ug}). If this is the case, these (or upgrades) should be able to determine the neutrino mass hierarchy. Otherwise, we will need to wait for ``next-next'' generation probes, which are sensitive to $\stheta$ down to $\sin^22\theta_{13} \sim 10^{-4}$
by, for instance, using the ``magic baseline'' at a neutrino factory~\cite{Huber:2003ak}. Again, if this turns out to be the case, these (or, say, a nearby supernova explosion) should be able to determine the neutrino mass hierarchy. If, on the other hand, $\theta_{13}$ is smaller still, we are ``stuck'' with detailed studies of muon neutrino disappearance \cite{deGouvea:2005hk}, and, perhaps, ``non-oscillation'' probes of neutrino masses (see, for example, \cite{Pascoli:2005zb,Choubey:2005rq,deGouvea:2005hj}).

In this study, we have re-visited the requirements for a neutrino mass hierarchy measurement
using $P_{\mu \mu}$, in the case $\stheta=0$. We need two very precise measurements of
the atmospheric oscillation frequency at very different values of $L/E$, one at a very
high oscillation maximum where solar effects contribute significantly (alternatively, a combination with antineutrino running could replace the combination
with a precise $\ldm$ measurement from a ``small'' $L/E$ value).
Theoretical estimates  (analysis at the probability level) point to the lowest attainable neutrino energies and very long baselines, $L\gtrsim 3000$~km.  In practice, on the other hand, the detector
energy resolution limits our ability to resolve fast oscillations (and hence ``measure'' $\Delta m^2_{31}$) and the optimal baseline is determined to be such that the largest number of events is obtained at relevant oscillation maxima. The conditions above require large event rates capable of overcoming the
large $1/L^2$ suppression of the beam flux with which we are required to deal, and excellent energy resolution in order to resolve the relevant low-energy oscillation maxima. We were especially interested in the whether ``current'' experimental setups meet these requirements, at least in principle. Not surprisingly, we find that, in order to significantly determine the character of the neutrino mass hierarchy, one is forced to push existing detector capabilities (and running times) to their technological  extremes.

Using conventionally established detector technology, we have demonstrated that a neutrino factory optimized for this purpose, operated at $L \simeq 6 \, 000 \, \mathrm{km}$ and with substantially increased statistics could do the mass hierarchy measurement for $\stheta=0$. From
the optimization,
 we find it best not to require charge identification in order to enhance the low energy detection
efficiencies, and we use a low muon energy $E_\mu=15 \, \mathrm{GeV}$ to relatively enhance
the flux in the low-energy region. For example, a TASD detector ten times as big as NOvA would allow, after $23$ years of running time, a $3 \sigma$ measurement
(for $10^{21}$ useful muon decays/year). Though these numbers are quite
large, note that the mass hierarchy measurement for $\stheta=0$ becomes most relevant if
no $\stheta$-signal is found. Assuming that substantial resources will go into this (assumed to be fruitless) search for $\theta_{13}$, it is reasonable to suppose that a significant amount of data and equipment (beams and detectors) will already be available
at the time we decide to measure the mass hierarchy as described here. In addition, the $6 \, 000 \, \mathrm{km}$ baseline may also be required
for a high-confidence level MSW~effect verification for $\stheta=0$ via the solar appearance
term~\cite{Winter:2004mt}.\footnote{Note that for this appearance channel measurement
a detector with charge identification is required. Therefore, one may want to
{\em a priori} choose a detector technology which can be magnetized at this baseline and analyze
the data set in two different samples (with and without charge identification).} The neutrino factory
measurement is limited by the flux-related
drop of events in the low energy region, which is quite common for
similar broad band beams. This means that in the region where the energy resolution limits
the performance, the event rates are already too low to have statistical significance, and
a better energy resolution would not help significantly.

A very promising alternative is a narrow band
off-axis superbeam similar to T2HK. Using a conventional water Cherenkov detector, the
effective energy resolution coming from Fermi motion of the nucleons limits the
ability to resolve higher oscillation maxima, and yields a performance much worse than that of 
the discussed neutrino factory. However, with modest improvements to the energy resolution, such a narrow band beam measurement combined with a measurement at a smaller $L/E$ (or
antineutrino running or a neutrino factory) proves to be significantly more sensitive to the neutrino mass hierarchy.
For example, if  the energy resolution were improved
by only a factor of two (from about $85 \, \mathrm{MeV}$ to about $43 \, \mathrm{MeV}$)
one would be able to determine the neutrino mass hierarchy  at the $90\%$~CL by combining eight years of data  from a T2HK-like experiment
at $L=4 \, 000 \, \mathrm{km}$ ($1 \, \mathrm{Mton}$ Water Cherenkov detector)  with data from a neutrino factory at $L = 6 \,000 \, \mathrm{km}$ ($50 \, \mathrm{kt}$ TASD). We note that the energy resolution limits the number of the
observable oscillation maxima and that it determines the optimum baseline for performing the measurements described here. For a fixed low-energy narrow band beam (peak energy below 1~GeV), we find that the optimal baseline increases (up to around $6 \, 000 \, \mathrm{km}$) as the energy resolution improves.

In summary, we have identified two important discriminators between the use of low-energy
narrow band beam technology and a relatively low energy neutrino factory: If $\ldm$
turned out to be substantially below the current best-fit value, or a detection technology
with a better minimum energy resolution than that of Water Cherenkov detectors
was identified, narrow band superbeams should be the appropriate choice for
a $\theta_{13}=0$ mass hierarchy measurement. If, on the other hand, $\ldm$ was
larger than the current best-fit value or no substantial improvement of the energy resolution
was obtained, a (``lower-energy,'' $E_{\mu}=15$~GeV) neutrino factory at about $6 \, 000 \, \mathrm{km}$ with a substantial
luminosity increase, or another beam with similar characteristics, could be the most efficient choice for the mass hierarchy measurement.

We conclude by reemphasizing that energy resolutions not far from the currently achievable would significantly
enhance the physics potential of this measurement. For example, different detector technologies and/or  detection materials (such as liquid argon)
could be used to improve the  energy resolution and our ability to determine the neutrino mass hierarchy --- provided that the ability to record events is comparable to the QE event rate in a $1 \, \mathrm{Mton}$ Water Cherenkov
detector. Since improving on the low-energy muon energy resolution has, so far, not been necessary
for the more conventional measurements, we wish to raise the level awareness of the community to  this issue.

\subsection*{Acknowledgments}

We would like to thank Dave Casper, Debbie Harris, Patrick Huber, Joe Sato, and Thomas Schwetz
for useful discussions and comments on various aspects of this study. We are especially thankful to Thomas Schwetz for comments on an earlier drafts of this manuscript.
The work of AdG is sponsored in part by the US Department of Energy Contract DE-FG02-91ER40684, while 
WW would like to
acknowledge support from the W.~M.~Keck Foundation and NSF grant PHY-0070928.

\end{document}